\documentclass[letter,twocolumn]{jpsj2} 
\title{Drastic Change of Magnetic Phase Diagram in Doped Quantum Antiferromagnet TlCu$_{1-x}$Mg$_x$Cl$_3$}

\author{Masashi \textsc{Fujisawa}$^{1}$\thanks{E-mail address: mercy@lee.phys.titech.ac.jp}, Toshio \textsc{Ono}$^{1}$, Hideyasu \textsc{Fujiwara}$^{1}$, Hidekazu \textsc{Tanaka}$^{2}$, Vadim \textsc{Sikolenko}$^{3}$, \\
Michael \textsc{Meissner}$^{3}$, Peter \textsc{Smeibidl}$^{3}$, Sebastian \textsc{Gerischer}$^{3}$ and Hans A. \textsc{Graf}$^{3}$}

\inst{$^{1}$Department of Physics, Tokyo Institute of Technology, Oh-okayama, Meguro-ku, Tokyo 152-8551 \\
$^{2}$Research Center for Low Temperature Physics, Tokyo Institute of Technology, Oh-okayama, Meguro-ku, Tokyo 15-8551 \\
$^{3}$Hahn-Meitner-Institute, Glienicker Starasse 100, D-14109 Berlin, Germany}

\abst{TlCuCl$_3$ is a coupled spin dimer system, which has a singlet ground state with an excitation gap of $\Delta/g\mu_{\mathrm B}$ = 5.5 T. TlCu$_{1-x}$Mg$_x$Cl$_3$ doped with nonmagnetic Mg$^{2+}$ ions undergoes impurity-induced magnetic ordering. Because triplet excitation with a finite gap still remains, this doped system can also undergo magnetic-field-induced magnetic ordering. By specific heat measurements and neutron scattering experiments under a magnetic field, we investigated the phase diagram in TlCu$_{1-x}$Mg$_x$Cl$_3$ with $x\sim 0.01$, and found that impurity- and field-induced ordered phases are the same. The gapped spin liquid state observed in pure TlCuCl$_3$ is completely wiped out by the small amount of doping.
 }

\kword{TlCuCl$_3$, TlCu$_{1-x}$Mg$_x$Cl$_3$, specific heat, neutron elastic scattering, magnetic-field-induced magnetic ordering, impurity-induced magnetic ordering, coupled spin dimer, spin gap}

\begin{document}
\maketitle
Recently, considerable attention has been paid to the quantum spin system with a gapped ground state. Such a spin gap system exhibits no magnetic ordering down to a zero temperature. Most gapped ground states arise from the dimerization of spins that are localized in magnetic ions. Hence, when nonmagnetic ions are substituted for magnetic ions, unpaired spins are produced, and staggered moments are induced around these unpaired spins. Unpaired spins can interact through effective exchange interactions that are mediated by spin dimers in between \cite{Sigrist,Imada,Yasuda}. Three-dimensional (3D) magnetic ordering can occur due to effective exchange interactions. Such impurity-induced magnetic ordering is observed in many spin gap systems \cite{Hase,Masuda1,Azuma,Uchiyama}.

This study is concerned with impurity- and field-induced magnetic orderings in the doped spin gap system TlCu$_{1-x}$Mg$_x$Cl$_3$. The parent compound TlCuCl$_3$ is a coupled spin dimer system, which has a singlet ground state with an excitation gap of $\Delta/k_{\rm B}$ = 7.5 K \cite{Takatsu,Oosawa1,Tanaka}. The small gap compared with the intradimer exchange interaction $J/k_{\rm B} = 65.9$ K is ascribed to strong interdimer exchange interactions \cite{Cavadini,Oosawa2}. In a magnetic field, the gap is suppressed and closes completely at the critical field $(g/2)H_{\rm c}=\Delta/g\mu_{\rm B} =5.5$ T. For $H > H_{\rm c}$, TlCuCl$_3$ undergoes magnetic ordering due to interdimer interactions \cite{Oosawa1}. Field-induced magnetic ordering in TlCuCl$_3$ has been extensively studied by various techniques including neutron scattering \cite{Tanaka,Rueegg}. The results obtained can be described as a result of the Bose-Einstein condensation of spin triplets (magnons)\cite{Nikuni,Matsumoto}. 

From a magnetization measurement and a neutron scattering experiment, it was found that TlCu$_{1-x}$Mg$_x$Cl$_3$ with $x\leq 0.03$ exhibits impurity-induced magnetic ordering \cite{Oosawa3,Oosawa4}. The spin structure of the impurity-induced phase is the same as that of the field-induced magnetic ordered phase for $H\parallel b$ in the parent compound TlCuCl$_3$. The easy axis lies in the $(0,1,0)$ plane at an angle of 13$^\circ$ from the $[2,0,1]$ direction to the $a$-axis. Spin flop transition was observed at $H_{\rm sf} \approx 0.3$ T, which is almost independent of $x$. Triplet excitation with a finite gap, which can be interpreted as excitation from intact dimers, was also observed \cite{Oosawa4}. With decreasing temperature, the gap increases below $T_{\rm N}$ in proportion to the order parameter. Mikeska {\it et al.} \cite{Mikeska} argued that the small staggered magnetic ordering induced in intact dimers gives rise to the enhancement in triplet gap.

Since the triplet gap still remains in TlCu$_{1-x}$Mg$_x$Cl$_3$, we can expect the occurrence of field-induced magnetic ordering. Although there are many theoretical and experimental studies of impurity-induced magnetic ordering in doped spin gap systems, the study of field-induced magnetic ordering in doped spin gap systems is limited. The relationship between impurity- and field-induced ordered phases has not been sufficiently understood. To investigate a magnetic phase diagram for temperature vs magnetic field in TlCu$_{1-x}$Mg$_x$Cl$_3$, we carried out a specific heat measurement and a neutron scattering experiment in magnetic fields.

Before preparing the doped TlCu$_{1-x}$Mg$_x$Cl$_3$ crystals, we prepared single crystals of TlCuCl$_3$ and TlMgCl$_3$. Mixing TlCuCl$_3$ and TlMgCl$_3$ in a ratio of $(1-x) : x$, we prepared TlCu$_{1-x}$Mg$_x$Cl$_3$ by the vertical Bridgman method. We obtained single crystals of $\sim 0.5$ cm$^3$. The magnesium concentration $x$ was analyzed by inductively coupled plasma$-$optical emission spectroscopy (ICP$-$OES) after the measurements. In the present study, we use samples for  with $x=0.0088$ and 0.015. The TlCu$_{1-x}$Mg$_x$Cl$_3$ crystal is cleaved easily parallel to the planes $(0,1,0)$ and $(1,0,\bar{2})$. 

Specific heat measurements were performed for the sample with $x = 0.0088$. Specific heat was measured down to 0.45 K at magnetic fields up to 9 T for $H \parallel b$, $H \perp (1,0,\bar{2})$ and $H \parallel [2,0,1]$, using a Physical Property Measurement System (Quantum Design PPMS) by the relaxation method. The same sample was used in the measurements for $H \perp (1,0,\bar{2})$ and $H \parallel [2,0,1]$, and another sample that was taken from the same batch was used in the measurement for $H \parallel b$. As will be shown later, no difference was observed between the ordering temperatures at a zero field obtained for both measurements. This implies a good homogeneity in the same batch.

Neutron scattering experiments were carried out using the E1 spectrometer of the BER II Research Reactor of the Hahn-Meitner Institute with the vertical field cryomagnet VM1. Incident neutron energy was fixed at $E_{\rm{i}} = 13.9$ meV, and the horizontal collimation sequence was chosen to be $40'-80'-40'-40'$. A single crystal with $x=0.015$ was used. The sample was mounted in the cryostat with its $(0, 1, 0)$ cleavage plane parallel to the scattering plane, so that reflections in the $a^*-c^*$ plane were investigated. The sample was cooled to 0.4 K using a $^3$He refrigerator. An external magnetic field of up to 12 T was applied along the $b$-axis. 

\begin{figure}[htbp]
  \begin{center}
    \includegraphics[keepaspectratio=true,width=85mm]{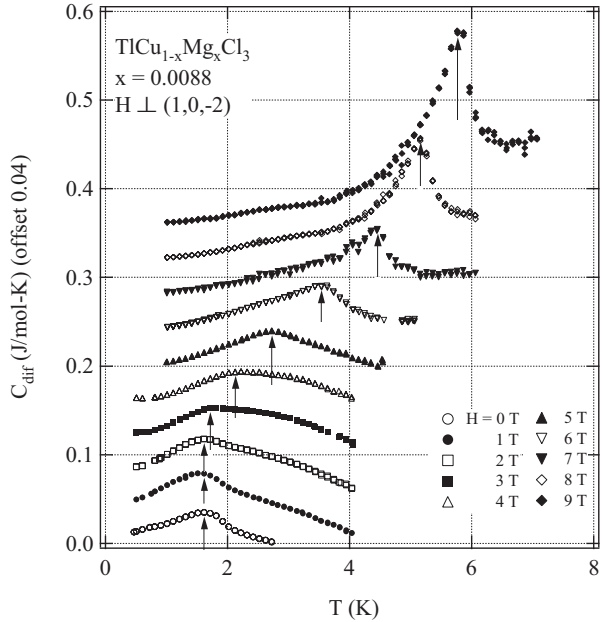}
  \end{center}
  \caption{Temperature dependence of $C_{\rm{dif}}$ ($=C(H)-C_0$) for $H\perp (1, 0, \bar{2})$, where $C (H)$ is the specific heat for TlCu$_{1-x}$Mg$_x$Cl$_3$ in a magnetic field $H$ and $C_0$ is the specific heat for TlCuCl$_3$ at $H = 0$. Arrows denote N\'{e}el temperatures.}
  \label{fig:hc.eps}
\end{figure}

\begin{figure}[htbp]
  \begin{center}
    \includegraphics[keepaspectratio=true,width=85mm]{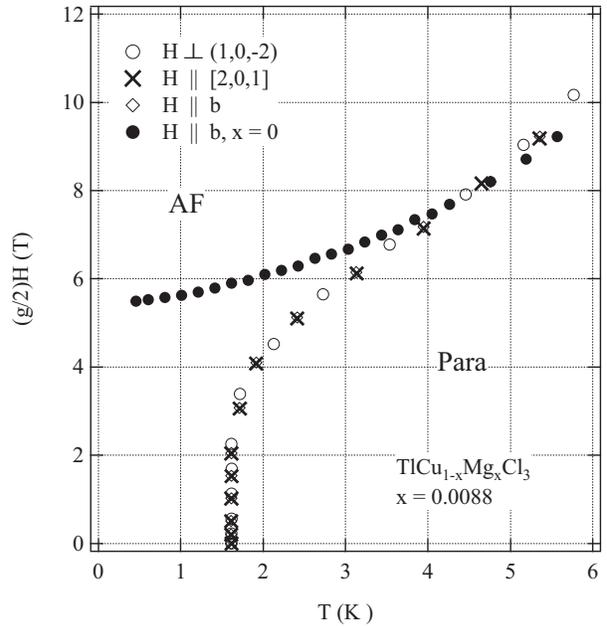}
  \end{center}
  \caption{Phase diagram for magnetic field vs temperature in TlCu$_{1-x}$Mg$_x$Cl$_3$. The magnetic field is applied for $H \parallel b$, $H \perp (1,0,{\bar 2})$ and $H \parallel [2,0,1]$.}
  \label{fig:pd.eps}
\end{figure}

Specific heat for TlCu$_{1-x}$Mg$_x$Cl$_3$ with $x = 0.0088$ displays a small cusplike anomaly below 6 K indicating magnetic phase transition at zero and finite magnetic fields. To determine the transition temperature $T_{\rm N}$ more accurately, we plotted the difference $C_{\rm {dif}}$ between the specific heat $C (H)$ for TlCu$_{1-x}$Mg$_x$Cl$_3$ in a magnetic field $H$ and $C_0$ for TlCuCl$_3$ at a zero field. Figure \ref{fig:hc.eps} shows the temperature dependence of $C_{\rm {dif}}$ for $H \perp (1,0, \bar{2})$ plane. The cusplike anomaly caused by phase transition is clearly observed. Arrows in Fig. \ref{fig:hc.eps} denote the transition temperatures.
The transition temperature at a zero field is $T_{\rm N}=1.7$ K. For $H\leq 3$ T, $T_{\rm N}$ is almost independent of temperature. However, for $H > 3$ T, $T_{\rm N}$ increases rapidly. A similar field dependence of $T_{\rm N}$ was also observed for $H\perp (1, 0, \bar{2})$ and $H\parallel [2,0,1]$. 
Figure \ref{fig:pd.eps} shows a summary of the data of $T_{\rm N}$ obtained at various magnetic fields for $x=0.088$ and those for TlCuCl$_3$. In Fig. \ref{fig:pd.eps}, the magnetic field is normalized by the $g$-factor. Here, we use $g=2.06, 2.23$ and 2.06 for $H \parallel b$, $H \perp (1,0,{\bar 2})$ and $H \parallel [2,0,1]$, respectively, which were determined for TlCuCl$_3$ by ESR \cite{Oosawa1}. We see that phase boundaries for these different field directions almost coincide. This fact indicates that the phase boundary is independent of the external field direction, when normalized by the $g$-factor, and that the anisotropy of the exchange interaction is negligible except in the low-field region ($H < 0.5$ T) including the spin flop field $H_{\rm sf}\approx 0.3$ T. 

With increasing external field, the ordering temperature $T_{\rm N}$ increases rapidly above $(g/2)H\simeq 3$ T. This field of $(g/2)H \sim$  3 T is considerably smaller than the gap field $(g/2)H_{\rm c} \sim$ 8 T corresponding to the triplet gap, which is estimated from the triplet gaps in TlCuCl$_3$ and TlCu$_{1-x}$Mg$_x$Cl$_3$ with $x\approx 0.03$ \cite{Oosawa4}. Since the triplet gap for intact dimers is enhanced with increasing doping rate $x$, the rapid increase in $T_{\rm N}$ for $H > 3$ T does not arise from the closing of the triplet gap for intact dimers. The effective exchange interaction $J_{\rm eff}$ between unpaired spins increases exponentially with decreasing triplet gap for intact dimers \cite{Sigrist,Imada}, which leads to an enhancement in ordering temperature. Thus, the rapid increase in $T_{\rm N}$ for $H > 3$ T should be attributed to the enhancement in $J_{\rm eff}$ due to the shrinkage of the triplet gap induced by the applied field. 

A notable feature of the magnetic phase diagram is that there is no boundary separating impurity- and field-induced antiferromagnetic phases, i.e., these two phases are contiguous. The gapped ground state that exists below $(g/2)H_{\rm c}=5.4$ T in pure TlCuCl$_3$ is completely wiped out by the small amount of nonmagnetic impurities.
The magnetic phase diagram for TlCu$_{1-x}$Mg$_x$Cl$_3$ is different from that of the doped Haldane system Pb(Ni$_{1-x}$Mg$_x$)$_2$V$_2$O$_8$ \cite{Masuda2}, in which impurity- and field-induced ordered phases are separated by a gapped disordered phase. 

\begin{figure}[htbp]
  \begin{center}
    \includegraphics[keepaspectratio=true,width=85mm]{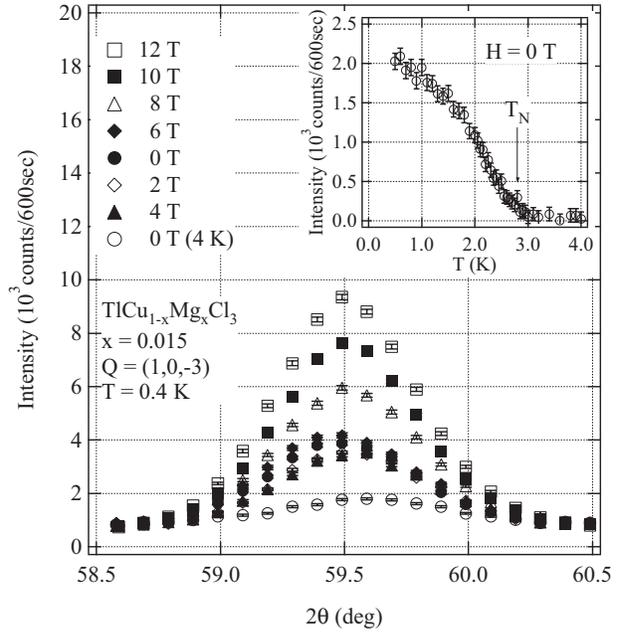}
  \end{center}
  \caption{Intensity of $\theta-2\theta$ scans at $T=0.4$ K for ${\mib Q} = (1, 0, \bar{3})$ reflection in TlCu$_{1-x}$Mg$_x$Cl$_3$ under various magnetic fields. The scan at the paramagnetic phase ($T=4.0$ K) is also plotted. Inset shows the temperature dependence of the magnetic peak intensity for ${\mib Q} = (1, 0, \bar{3})$, where temperature independent background is subtracted.}
  \label{fig:profile_10i3.eps}
\end{figure}

If the impurity- and field-induced ordered phases are identical, then the order parameters should be the same. Therefore, we can expect that the sublattice magnetization corresponding to the order parameter exhibits unusual field and temperature dependences. Then, we performed a neutron elastic scattering experiment on TlCu$_{1-x}$Mg$_x$Cl$_3$ with $x=0.015$, for which impurity-induced magnetic ordering occurs at $T_{\rm N}=2.8$ K as shown in the inset of Fig. \ref{fig:profile_10i3.eps}. Magnetic reflections with a resolution-limited width were observed at $\mib Q = (h, 0, l)$ for an integer $h$ and an odd $l$. These reciprocal points are the same as those for field- and pressure-induced transverse N\'{e}el orderings in TlCuCl$_3$ \cite{Tanaka,Oosawa5}. Figure \ref{fig:profile_10i3.eps} shows ${\theta}-2{\theta}$ scans for the $(1, 0, \bar{3})$ reflection measured at $T=0.4 $ K ($< T_{\rm N}$) under the various magnetic fields. We also plotted in Fig. \ref{fig:profile_10i3.eps} the same scan for $H=0$ at $T = 4.0$ K ($> T_{\rm N}$). Since the parent compound TlCuCl$_3$ belongs to the space group $P2_1/c$, nuclear peaks are expected only at $\mib Q = (h, 0, l)$ with an even $l$. However, as shown in Fig. \ref{fig:profile_10i3.eps}, weak nuclear peaks are observed for an odd $l$. Since no structural phase transition was observed in TlCu$_{1-x}$Mg$_x$Cl$_3$ through magnetization and specific heat measurements, we infer that the nuclear peaks observed at $\mib Q = (h, 0, l)$ with an odd $l$ are due to the local disturbance of the lattice caused by Mg$^{2+}$ doping.  

\begin{figure}[htbp]
  \begin{center}
    \includegraphics[keepaspectratio=true,width=85mm]{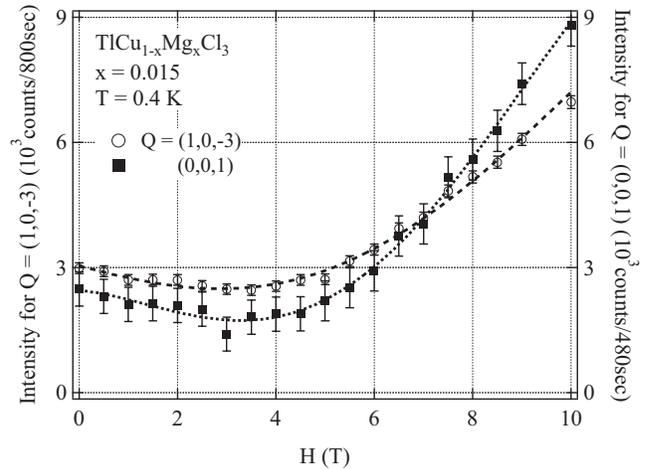}
  \end{center}
  \caption{Magnetic field dependence of magnetic Bragg peak intensities for ${\mib Q} = (1, 0, \bar{3})$ and $(0, 0, 1)$ reflections in TlCu$_{1-x}$Mg$_x$Cl$_3$. Dotted and broken lines are visual guides.}
  \label{fig:fieldscan.eps}
\end{figure}

As shown in Fig. \ref{fig:profile_10i3.eps}, magnetic Bragg intensity decreases slightly with increasing external field up to 4 T. For $H\geq 6$ T, the intensity increases rapidly. We investigated the field dependence of magnetic Bragg intensity in detail. Figure \ref{fig:fieldscan.eps} shows the intensity vs magnetic field plots for $\mib Q=(1,0,\bar{3})$ and $(0,0,1)$ magnetic peaks. Nuclear Bragg intensities and background are subtracted. With increasing external field, the intensities of both magnetic peaks have a minimum at $H\sim 3.5$ T and increase rapidly. However, no anomaly indicative of field-induced phase transition is observed. This result confirms that there is no boundary separating impurity- and field-induced ordered phases. Since Bragg peak intensities for both $\mib Q=(1,0,\bar{3})$ and $(0,0,1)$ exhibit similar field dependences, these field dependences are attributed not to the change in spin direction, but to the change in the magnitude of the ordered moment. The present result indicates that impurity- and field-induced ordered phases in TlCu$_{1-x}$Mg$_x$Cl$_3$ are the same, and that the order parameter has a minimum at $H\simeq 3.5$ T. 

\begin{figure}[htbp]
  \begin{center}
    \includegraphics[keepaspectratio=true,height=50mm]{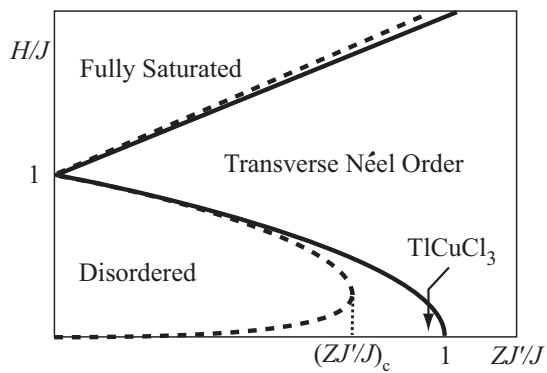}
  \end{center}
  \caption{Phase diagram of ground state proposed by Mikeska \textit{et al}\cite{Mikeska}. Solid and dashed lines are boundaries for the pure and doped systems. Arrow denotes the ($ZJ'/J$) corresponding to TlCuCl$_3$.}
  \label{fig:mikeska.eps}
\end{figure}

Recently, Mikeska {\it et al.} have theoretically investigated the interplay of the doping and the external field in the coupled $S=1/2$ Heisenberg spin dimer system on the basis of the mean-field approximation \cite{Mikeska}. For the ground state, they presented the schematic phase diagram in the $(ZJ'/J, H/J)$ plane as shown in Fig. \ref{fig:mikeska.eps}, where $J$ and $J'$ are the intradimer and interdimer interactions, respectively, and $Z$ is the coordination number. Their results are summarized as follows: 
doping causes the antiferromagnetic order at a zero field. For smaller $(ZJ'/J)$, there are three critical fields $H_{\rm{c1}}$, $H_{\rm{c2}}$ and $H_{\rm s}$. For $H < H_{\rm{c1}}$, the unpaired spins form an antiferromagnetic order through the effective exchange interaction $J_{\rm eff}$, and a small transverse staggered order also occurs in intact dimers. Unpaired spins are fully polarized at $H = H_{\rm{c1}}$, and the small transverse staggered order in intact dimers vanishes. The critical field $H_{\rm{c1}}$ is proportional to the impurity concentration $x$. Above $H_{\rm{c1}}$, the ground state is gapped and disordered. At the second critical field $H_{\rm{c2}}$, the triplet gap for intact dimers closes and the transverse staggered order occurs again in intact dimers.  With a further increase in magnetic field, the saturation takes place at $H=H_{\rm s}$. On the other hand, for larger $(ZJ'/J)$, the field range of the disordered state shrinks. There is a critical value $(ZJ'/J)_{\rm c}$, above which impurity- and field-induced ordered states merge. The point given by $(ZJ'/J) = 1$ is the quantum critical point that separates the gapped disordered state and the antiferromagnetic ordered state in the pure system. 

For TlCuCl$_3$, the critical field corresponding to the gap and saturation field are $(g/2)H_{\rm c}$ = 5.5 T and $(g/2)H_{\rm s} \sim$ 90 T, respectively. Hence, the exchange parameter should be located at the point indicated by the arrow in Fig. \ref{fig:mikeska.eps}. $(ZJ'/J)$ corresponding to TlCuCl$_3$ is close to unity and should be larger than $(ZJ'/J)_{\rm c}$ for $x\geq 0.0088$. In this parameter region, the field-induced disordered state is absent. The present experimental result that the impurity- and field-induced ordered states are connected without a boundary in TlCu$_{1-x}$Mg$_x$Cl$_3$ is in accordance with the theory by Mikeska {\it et al} \cite{Mikeska}. In the present system, the interdimer interaction is fairly large so that the triplet gap for intact dimers closes in a magnetic field before unpaired spins are fully polarized. For this reason, the impurity- and field-induced ordered states merge. 

According to the mean-field theory by Mikeska {\it et al} \cite{Mikeska}, $(ZJ'/J)_{\rm c}$ increases with decreasing impurity concentration $x$, and approaches unity for $x\rightarrow 0$. However, it is also considered that for small $x$, there is a finite critical value $(ZJ'/J)_{\rm c}$ being smaller than unity, because field-induced phenomena at $T=0$ should be determined by the local interactions between unpaired spin and induced moments around the unpaired spin that are independent of $x$ for $x\rightarrow 0$. Thus, it is interesting to investigate whether field-induced disordered state is realized in TlCu$_{1-x}$Mg$_x$Cl$_3$ with further decreasing $x$. 

In conclusion, we have performed specific heat measurements and neutron elastic scattering experiments on the doped spin gap system TlCu$_{1-x}$Mg$_{x}$Cl$_3$ with $x\sim 0.01$ in magnetic fields to investigate the relation between impurity- and field-induced ordered phases. Well-defined phase transitions were observed. It was found that the impurity- and field-induced ordered phases are contiguous, i.e., these two phases are the same. The gapped spin liquid state observed in pure TlCuCl$_3$ is completely wiped out by the small amount of doping. This is because the interdimer exchange interaction $J'$ is so strong in TlCu$_{1-x}$Mg$_{x}$Cl$_3$ that the triplet gap in intact dimers closes before unpaired spins near impurities are fully polarized by the external field. From the neutron scattering result, we found that the magnitude of the ordered moment has a minimum at $H\simeq 3.5$ T. It is considered that this unique behavior of the order parameter results from the competition between magnetic ordering and spin gap, the latter of which acts to separate impurity- and field-induced phases.

The authors would like to thank A. Oosawa and H. -J. Mikeska for stimulating discussions. They are also grateful to H. Imamura for his technical support.
This work was supported by a Grant-in-Aid for Scientific Research and the 21st Century COE Program at Tokyo Tech ``Nanometer-Scale Quantum Physics'' both from the Ministry of Education, Culture, Sports, Science and Technology of Japan.

\end{document}